\documentclass[aps,preprint,superscriptaddress, amsmath,amssymb,floatfix,longbibliography]{revtex4-1} 
\usepackage{bm, comment, graphicx, here, amsmath, color, mathrsfs}
\usepackage[normalem]{ulem}
\usepackage[colorlinks=True, allcolors=blue]{hyperref}
\usepackage{balance, setspace, float, titlesec, array}
\usepackage{ulem}
\titlespacing*{\section}{0pt}{1ex}{0ex}
\titlespacing*{\subsection}{0pt}{1ex}{0ex}

\newcommand{\blue}[1]{ \textcolor{blue}{#1} }
\newcommand{\red}[1]{ \textcolor{black}{#1} }

\usepackage{soul}
\soulregister\cite7
\soulregister\ref7
\soulregister\onlinecite7

\begin{document}
\title{Constructing material network representations for \\intelligent amorphous alloys design}

\author{S.-Y. Zhang$^{\dagger}$}
\affiliation{Songshan Lake Materials Laboratory, Dongguan, China}

\author{J. Tian$^{\dagger}$}
\affiliation{Lawrence Berkeley National Laboratory, Berkeley CA 94720, USA}

\author{S.-L. Liu$^{\dagger}$}
\affiliation{Songshan Lake Materials Laboratory, Dongguan, China}

\author{H.-M. Zhang}
\affiliation{Songshan Lake Materials Laboratory, Dongguan, China}

\author{H.-Y. Bai}
\affiliation{Songshan Lake Materials Laboratory, Dongguan, China}
\affiliation{Institute of Physics, Chinese Academy of Sciences, Beijing, China}

\author{Y.-C. Hu}
\email{Email: yuanchao.hu@sslab.org.cn \\
	$^\ddagger$These authors contributed equally: S. Z, J. T, S. L.}
\affiliation{Songshan Lake Materials Laboratory, Dongguan, China}

\author{W.-H. Wang}
\affiliation{Songshan Lake Materials Laboratory, Dongguan, China}
\affiliation{Institute of Physics, Chinese Academy of Sciences, Beijing, China}

\date{May 15, 2025}

\vspace{5mm}

\begin{abstract}
Designing high-performance amorphous alloys is demanding for various applications. But this process intensively relies on empirical laws and unlimited attempts. The high-cost and low-efficiency nature of the traditional strategies prevents effective sampling in the enormous material space. Here, we propose material networks to accelerate the discovery of binary and ternary amorphous alloys. The network topologies reveal hidden material candidates that were obscured by traditional tabular data representations. By scrutinizing the amorphous alloys synthesized in different years, we construct dynamical material networks to track the history of the alloy discovery. We find that some innovative materials designed in the past were encoded in the networks, demonstrating their predictive power in guiding new alloy design.  These material networks show physical similarities with several real-world networks in our daily lives. Our findings pave a new way for intelligent materials design, especially for complex alloys.
\end{abstract}

\maketitle

\section*{Introduction}

Advancing the fundamental disciplines, such as materials science, physics, and chemistry, to name a few, intensively relies on the discovery of new materials~\cite{de_materials_2021,merchant_scaling_2023}. The successful fabrication of material in experiments quite often signifies the opening of an unknown world. It also usually serves as a carrier for the acquisition of theoretical knowledge to catalyze next-generation optimization. Iterating this process presents the common evolution pathway for critical scientific breakthroughs in natural science.

From the periodic table, there is only a limited number of elements so far. Nevertheless, the enormous way of optimal combinations among these elements makes up the Earth, enabling persistent lives in different biological groups. It is fair to say that the material space is infinite~\cite{merchant_scaling_2023,li_how_2017}. Ever since ancient times, materials were developed mainly by trial-and-error experiments, with empirical laws summarized from mostly the failure experiences. 
In modern times, there is undoubtedly high demand for materials with superior performances. Unfortunately, the traditional strategy fails to satisfy this requirement. 

To circumvent this grand challenge, various methodologies have been proposed in the past. For example, high-throughput sputtering experiments are capable of synthesizing a library of more than 1,000 ternary alloys with composition gradients in just one experiment~\cite{ding_combinatorial_2014, li_high_2019, li_data_2022}. 
Meanwhile, motivated by the soaring computational power, parallel computer simulations by either density functional theory or molecular dynamics provide considerable theoretical support for experimental material design~\cite{jain_commentary_2013,batra_emerging_2021,curtarolo_NM_2013}.
With the accumulated scientific data, advanced machine learning techniques have witnessed their fast growth in the domain of materials prediction~\cite{merchant_scaling_2023,batra_emerging_2021}.
In particular, the advent of artificial intelligence (AI) in computer vision and natural language processing greatly fostered the virtual screening of candidate materials in the large latent space~\cite{szymanski_autonomous_2023, liu2024prompt, bran_augment_2024}.

However, several key problems remain to accelerate materials discovery. The most critical one is the severe shortage of high-quality data. This is inherent in the complex and costly process of material synthesis. Only successful trials were recorded in the literature, as well.

One prominent example is amorphous alloys, which are mainly made of transition metals but lack long-range translational periodicity. The infinite possibilities of atomic packing in space at the microscopic scale puts further difficulty in tackling this problem. Since its first discovery in 1960~\cite{klement1_non_1960}, several decades passed by but glassiness has been successfully seen in only $\sim670$ systems, ranging from binary to multi-component alloys over about 7,000 compositions~\cite{ward_general-purpose_2016}. The largest sample size is still limited to several centimeters, but the general size is around millimeters and even smaller~\cite{takeuchi_classification_2005}. These situations remarkably constrain the applicable services of amorphous alloys in various fields, benefiting from their metal-like mechanics and glass-like functionalities. There are ongoing efforts to address this problem by high-throughput experiments~\cite{ding_combinatorial_2014, li_high_2019} and supervised machine learning over experimental data~\cite{ward_general-purpose_2016,sun2017machine,xiong2020machine,liu2020machine,zhou_rational_2021, afflerbach_machine_2022,yao2022balancing,liu2023machine, forrest2023evolutionary, zhou2023generative,liu2024effective}, or their combinations~\cite{ren_accelerated_2018, li_data_2022}. However, the candidate materials are far from fulfilling the demands. 

To speed up this process, the emerging data-driven techniques are promising, with possibly the assistance of cutting-edge AI technologies~\cite{merchant_scaling_2023,szymanski_autonomous_2023}. As the amorphous alloys were generated by immense research efforts, there is tremendous hidden physics behind the small amount of the available dataset~\cite{karniadakis_physics_2021}. 
How to design effective materials representation is a critical step in data mining, no matter for researchers or machines.
To learn from these data, previous studies used the physical properties of the alloys and their elements to create learnable features for machines in a supervised manner~\cite{ward_general-purpose_2016,sun2017machine,xiong2020machine,liu2020machine,zhou_rational_2021, afflerbach_machine_2022,yao2022balancing,liu2023machine, forrest2023evolutionary, zhou2023generative,liu2024effective}. A tabular dataset is structured by using alloy compositions as rows while their features as columns. These instances are treated independently during learning. While this supervised learning provides some insights into inventing new amorphous alloys, the prediction power is still limited. One of the crucial reasons is the plain representation in the tabular form. The hidden physical connections among alloys are neglected. The information loss seriously deteriorates the model predictability. Physics-inspired data mining will be mandatory to build excellent machine-learning models. Thus, a more sophisticated data representation, especially for a small amount of data with lots of physics, is critical for successful data learning to assist the optimal design of amorphous alloys.

In this work, we propose an effective network representation for amorphous alloys. We focus on the binary and ternary alloys that are prone to form glass by rapid-quenching experiments. The datasets are collected from the joint efforts over the past decades in the literature~\cite{ren_accelerated_2018, ward_machine_2018, li_how_2017,schultz_exploration_2021}. We construct both virtual and realistic networks from these data. They provide valuable scientific senses to better understand the materials and the associated hidden physics. By thorough network mining, we identify the possible material space and the evolution pathway of amorphous alloy development. The imbalanced contributions from various elements are revealed. Furthermore, by analyzing the dynamical material networks built in different years, the history of the alloy discovery is unveiled. 
We find that some innovative materials fabricated in the past were actually encoded in the material networks. This demonstrates the predictive power of these networks.
The topology analysis provides unique perspectives to further optimize the material design procedure. 
In addition, from the network analysis, these networks show features belonging to the abnormal scale-free class, which exhibits preferential attachment in the network growth. We argue that these material networks intrinsically suffer from physical constraints from the available constituent elements in the periodic table, which distinguishes them from the general scale-free class~\cite{barabasi_emergence_1999}. 
We find similar features in several real-world networks in our daily lives.
This bridges material networks to realistic ones, facilitating collaborative research from multiple disciplines and paving new avenues to intelligently discovering complex alloys.

\vspace{1cm}
\section*{Results}\label{sec2}
\subsection*{Material networks of amorphous alloys}

\begin{figure}[!t] 
	\centering
	\includegraphics[width = \textwidth]{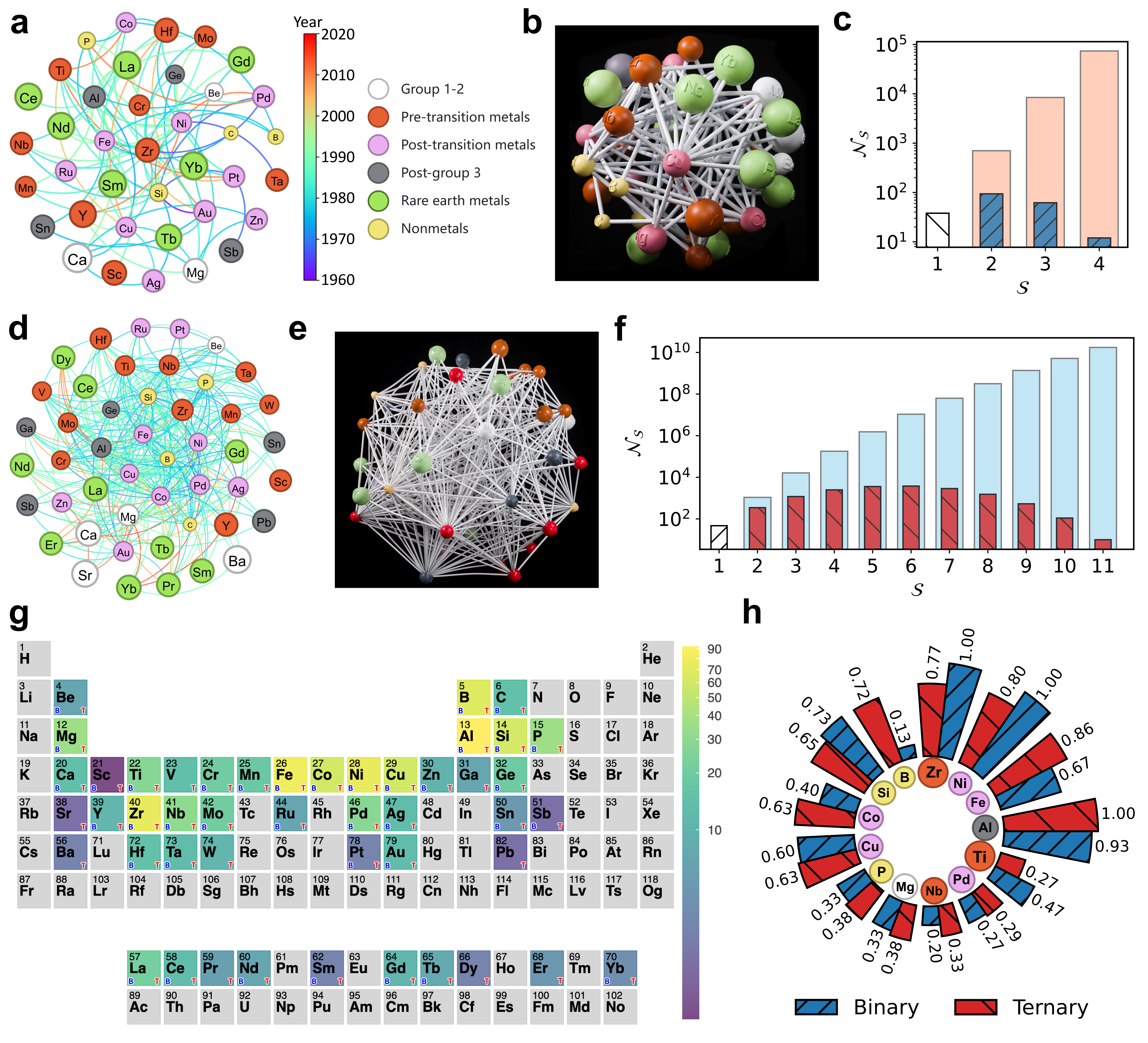}
	\caption{
		{\bf Network representations and analyses for amorphous alloys.}
		(a) Network of binary alloys. The nodes are different elements colored by their periodic groups. Their diameters are scaled based on the atomic radii. Each edge dictates an experimentally discovered amorphous alloy with the two nodes at a certain year (see color bar). There are 38 nodes and 94 edges. 
		(b) The network of binary alloys by 3D printing.
		(c)  Clique size distribution of the binary network. $\mathcal{N}_{\mathcal{S}}$ of $\mathcal{S}=1$ gives the number of nodes in (a). The light-shaded bars represent all possible geometric objects (${\bf C}_{38}^{\mathcal{S}}$) in the network, like rectangles for $\mathcal{S}=4$.
		(d) Network of ternary alloys. It is plotted in the same way as (a) but based on triangles. Each triangle represents a ternary system. There are 47 nodes and 352 triangles, generating 348 edges.
		(e) The network of ternary alloys by 3D printing.
		The color schemes in (d) and (e) are the same as (a) and (b), respectively.
		(f) Clique size distribution of the ternary network, similar to (c). The numbers for the light-shaded bars are ${\bf C}_{47}^{\mathcal{S}}$.
		(g) A periodic table highlighting elements used in (d). The color bar depicts the number of ternary alloys an element is involved in. The big letter ``B" or ``T" marks an element showing up in the binary or ternary network.
		(h) The numbers of edges and triangles of an element in the binary and ternary networks, normalized by the corresponding largest number. The largest degree in (a) is 16, and the largest number in (g) is 93.
	}
	\label{fig1}
\end{figure}

We start our investigation by carefully collecting the available data in the literature. (see \blue{Materials and Methods})
As the data are widely dispersed in various journals, we may not have all the data reported so far but should have the vast majority.
Comprehensive data cleanings, such as redundancy removal, composition correction, and quality alignment, are performed to create a high-quality database by combining the existing datasets~\cite{ren_accelerated_2018, ward_machine_2018, li_how_2017,schultz_exploration_2021}. This process is essential and straightforward but requires intensive efforts with carefulness. It is of paramount importance in determining the research quality afterward. 
In addition, other than the common data information used in the previous studies, we scrutinize the relevant publications to extract the earliest reporting year for an alloy composition. This helps perform a dynamic analysis of the materials data. Dynamical material networks are thus generated.
At the current stage, we neglect the specific alloy composition but only focus on the alloy system. 
It is beneficial to avoid unnoticed uncertainties and undesired parsing errors. Meanwhile, this strategy permits better tolerance for data analysis and future model prediction. We aim to learn well from this coarse-graining strategy before targeting the ambition of pinpointing a specific alloy composition. 
This also distinguishes us from the previous works.

We separate the database into two groups based on the number of constituents, i.e. binary and ternary alloys. Reports on the alloys with more components are comparatively rare.
For the binary alloys, there are 94 systems with 38 elements. While there are 352 ternary systems with 47 elements. Based on the fundamental entities, being an edge or a triangle, we construct the material networks for the binary and ternary alloys, respectively. 
That is, with the 38 elements as nodes, we add a link between two elements if there is a binary alloy that can form a metallic glass, i.e. an existing data point in the database.
For the ternary systems, we add a triangle with three links if there is a ternary alloy that can form a metallic glass. 
We name them binary network and ternary network below for consistent terminology. 
Thus, we stress that all our following data analysis is based on edges for the binary network and on triangles for the ternary network, even though the basic components of a graph are nodes ($0d$, $d$: dimensionality) and edges ($1d$). A triangle ($2d$) is a high-order object, which is important in shaping topology dynamics~\cite{millan_topology_2025}.

The spatial layouts of these networks are optimized by using the Fruchterman-Reingold algorithm, as shown in Fig.~\ref{fig1}a for the binary network and Fig.~\ref{fig1}d for the ternary network. (see \blue{Materials and Methods}) 
To enrich the visualization, the nodes are sized by the metallic radii of the elements, which are then colored by their positions in the periodic table. Meanwhile, the edges are colored by the earliest invention year of the alloy system over different compositions. This strategy is utilized for both the binary network and the ternary network. 
Because of the large number of links in these networks, especially the ternary, they are too fuzzy to provide clear insights vividly. To solve this problem, we come up with the idea of building the corresponding realistic networks by 3D printing. Figure~\ref{fig1}b and ~\ref{fig1}e illustrate the printed networks by carefully adjusting the layouts so that they are rigid enough. This is neglected in computation but reminds the importance of layout optimization~\cite{boguna_network_2021}.
Usually the concept ``network" refers to a realistic object while ``graph" is virtual, but here we mix them for easier discussion. These networks provide valuable three-dimensional insights to inspire our data analysis and graph mining in computation.

Motivated by these material networks, we begin to explore the material space. We first count the existing cliques in the networks. A clique is defined as a geometric object with each pair of its nodes connected. The number of nodes in a clique gives its size $\mathcal{S}$. 
For instance, a triangle has $\mathcal{S}=3$. In principle, $\mathcal{S}$ is mapped to the number of components of an alloy, but in a strict manner. 

Figure~\ref{fig1}c depicts the count distribution of the cliques in the binary network. The number of cliques $\mathcal{N}_\mathcal{S}$ with $\mathcal{S}=2$ in the binary network is 94.
In fact, there are higher order objects forming automatically, even though only data with $\mathcal{S}=2$ are actually used. For example, there are fully connected triangles and rectangles existing, suggesting three-component and four-component metallic glass-formers. Furthermore, by directly considering all combinations of different numbers of elements (${\bf C}_{38}^\mathcal{S}$), we include the corresponding numbers in Fig.~\ref{fig1}c for comparison. Apparently, with more components, the alloy space increases exponentially with enlarged gaps from cliques.
These demonstrate rising challenges to discovering a multi-component system, let alone a specific alloy composition. 
The contrast in this plot also suggests the first-worthy-trial candidates in the largely unexplored material space inspired by the binary network, which would be overlooked by conventional tabular data representations.

These analyses apply to the ternary network, as shown in Fig.~\ref{fig1}f. Surprisingly, there are even more higher-order cliques forming automatically, with $\mathcal{S}$ up to 11. $\mathcal{N}_{\mathcal{S}}$ for $\mathcal{S}=1$ and $\mathcal{S}=3$ gives the numbers of elements and ternary alloys, respectively. The other quantities depict the fully connected multi-component systems in the current network. Although the number of all possible combinations (${\bf C}_{47}^\mathcal{S}$) increases dramatically from $\sim10^3$ to $\sim10^{11}$ with ascending components, $\mathcal{N}_{\mathcal{S}}$ of cliques behaves like a normal distribution peaking at $\mathcal{S}=5$ or 6. Thus, the ternary network conveys more messages and provides alloy design advice for alloys with a wide range of components. Investigating these candidates can by itself serve as the first experimental advice. It also emphasizes the limited space that the human beings have ever explored. This power is naturally missed in the traditional tabular representations. The difficulty of pinpointing a specific multi-component alloy aligns with the efforts to uncover a hidden star in the Universe. We leave this challenge of quantitatively unveiling the relationships between these groups and with experimental discoveries in the future studies. 

In addition, the elemental contributions to the networks are justified.
In Fig.~\ref{fig1}g, we impose the involved elements in metallic glass-formers in the periodic table. As expected, most of the elements are transition metals with rare-earth elements included. Some metalloids usually appear in amorphous alloys by minor alloying. The elements in the ternary network consist of those in the binary network, suggesting their hidden correlation. There are many more elements to be tested in future materials design. More quantitatively, the top 13 elements that most frequently make an amorphous alloy are illustrated in Fig.~\ref{fig1}h for the ternary network, compared to the binary one. While Al appears in most of the ternary alloys, Ni and Zr are the most for binary alloys. They can be defined as elemental hubs in the material networks. (see below) This comparison indicates the unbalanced contribution of these elements in making up the material networks, implying limited knowledge or preferential attachment~\cite{albert_statistical_2002}. 

\begin{figure}[!t] 
	\centering
	\includegraphics[width = 0.8\textwidth]{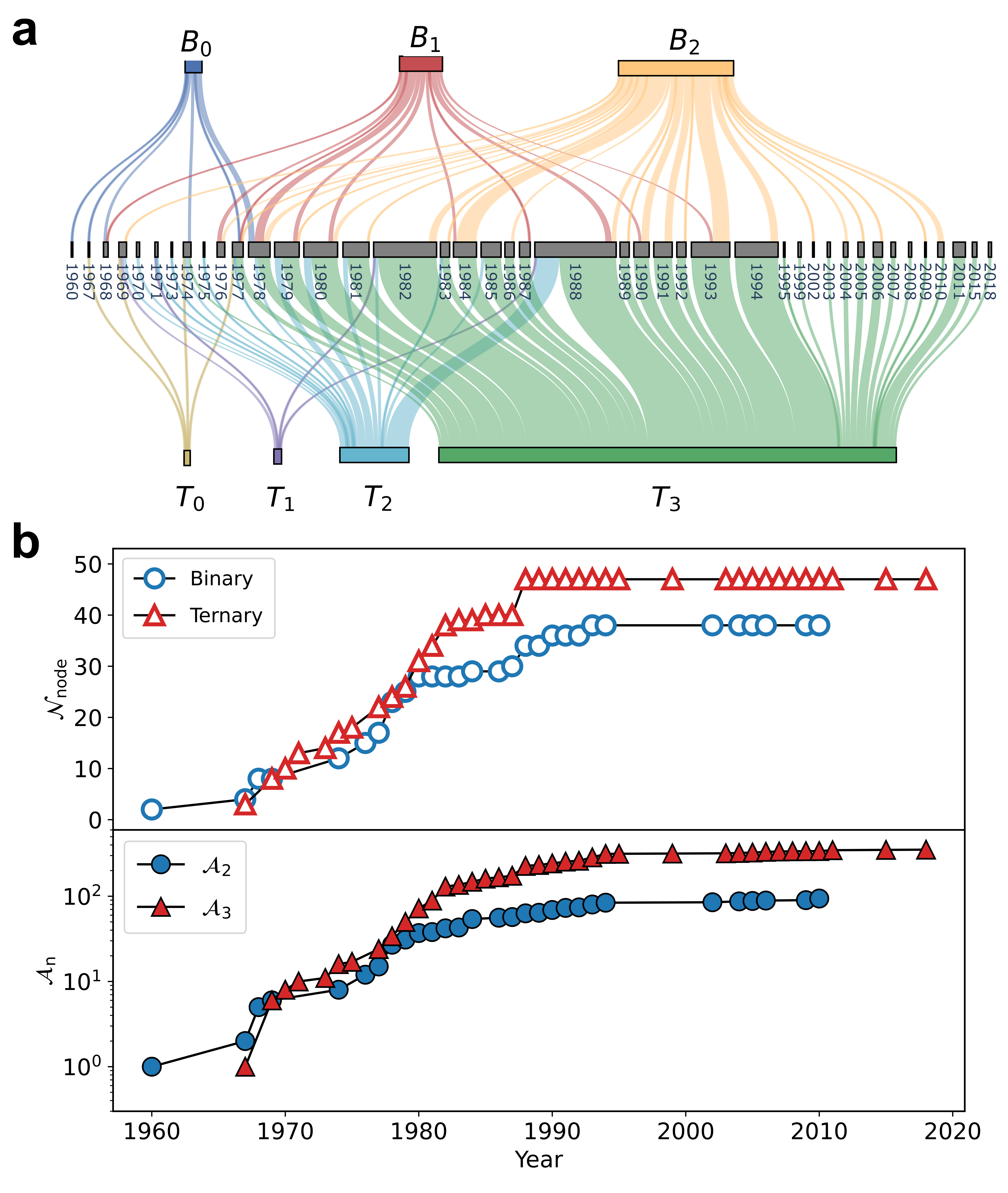}
	\caption{
		{\bf Discovery history of amorphous alloys.}
		(a) A Sankey diagram illustrating the invention histories of binary ($B_n$) and ternary ($T_n$) amorphous alloys from 1960. They are classified based on how many elements have been used in the previously developed amorphous alloys. For example, $B_1$ of 1988 demonstrates that one of the elements was used in the binary alloys developed before 1988.
		(b) Topology evolution of dynamical networks. The upper panel shows the number of nodes $\mathcal{N}_{\rm node}$, while the lower panel exhibits the number of links ($\mathcal{A}_2$) and triangles ($\mathcal{A}_3$) for the binary and ternary network, respectively.
	}
	\label{fig2}
\end{figure}

\subsection*{Dynamical material networks}

After recognizing the strength of network representation, we take the invention year of each system into consideration and construct the dynamical material networks by using the accumulated data to compensate data scarcity. In greater detail, the network of 1988 is constructed by all the nodes and edges/triangles discovered before 1988 and in 1988. This will help track the evolution path of amorphous alloys.
Figure~\ref{fig2}a demonstrates a Sankey diagram of the history of glass discovery in the binary network and the ternary network. Each set ($B_n$ or $T_n$) is further classified into sub-groups with different numbers of elements ($n$) that have been seen in previously invented alloys. For instance, $B_0$ and $T_0$ are brand new alloy systems with the involved elements never being touched before. $T_3$ dictates a class where all of the elements have been utilized before in other alloys but not in the new one. In the early stages, mainly before 1978, the brand-new alloy systems dominate, demonstrating highly creative exploration. Later on, there are more and more systems developed based on the used elements in both networks. Especially after 1988, nearly no new element is employed anymore. This means that the nodes already appear in the network, but either a link or a triangle is missing. This verifies the effectiveness of the material candidates from the cliques in Figs.~\ref{fig1}c, f.

This material invention pattern is further quantitatively proved by the yearly growth of the number of nodes $\mathcal{N}_{\rm node}$ (upper panel) and the number of entities $\mathcal{A}_n$ (link or triangle, lower panel) from the dynamical networks in Fig.~\ref{fig2}b. Especially, $\mathcal{N}_{\rm node}$ saturates after nearly 1990. The period of 1970-1980 witnessed a fast growth of $\mathcal{N}_{\rm node}$ and $\mathcal{A}_n$, giving rise to a plethora of new alloys that fertilize the blossom of the field development. After 2000, the alloy inventory enters into a suspending state. Investigating various properties of these amorphous alloys has become a main research focus since then.

These features highlight the effectiveness of the network representation, particularly in the dynamical form. From the opposite perspective, there emerges an innovation trap in the materials design, inevitably after many elements are tested. The knowledge acquisition from the literature motivates newcomers to explore the hidden connections between existing elements in the invention pool, not truly randomly. 
The designed materials may also already exist in the constructed material networks. (see more discussion below)
In essence, a knowledge graph was constructed individually that guides personal alloy design, which, in some sense, limits the research creativity. 
We may expect this innovation trap in many other materials fields as well, which is intriguing to explore in the future.
It thrills us that involving new elements may bring unexpected breakthroughs~\cite{li_high_2019}. Both enriching and enlarging the networks are crucial for future design.

\subsection*{Topology analysis of material networks}

\begin{figure}[!t] 
	\centering
	\includegraphics[width = \textwidth]{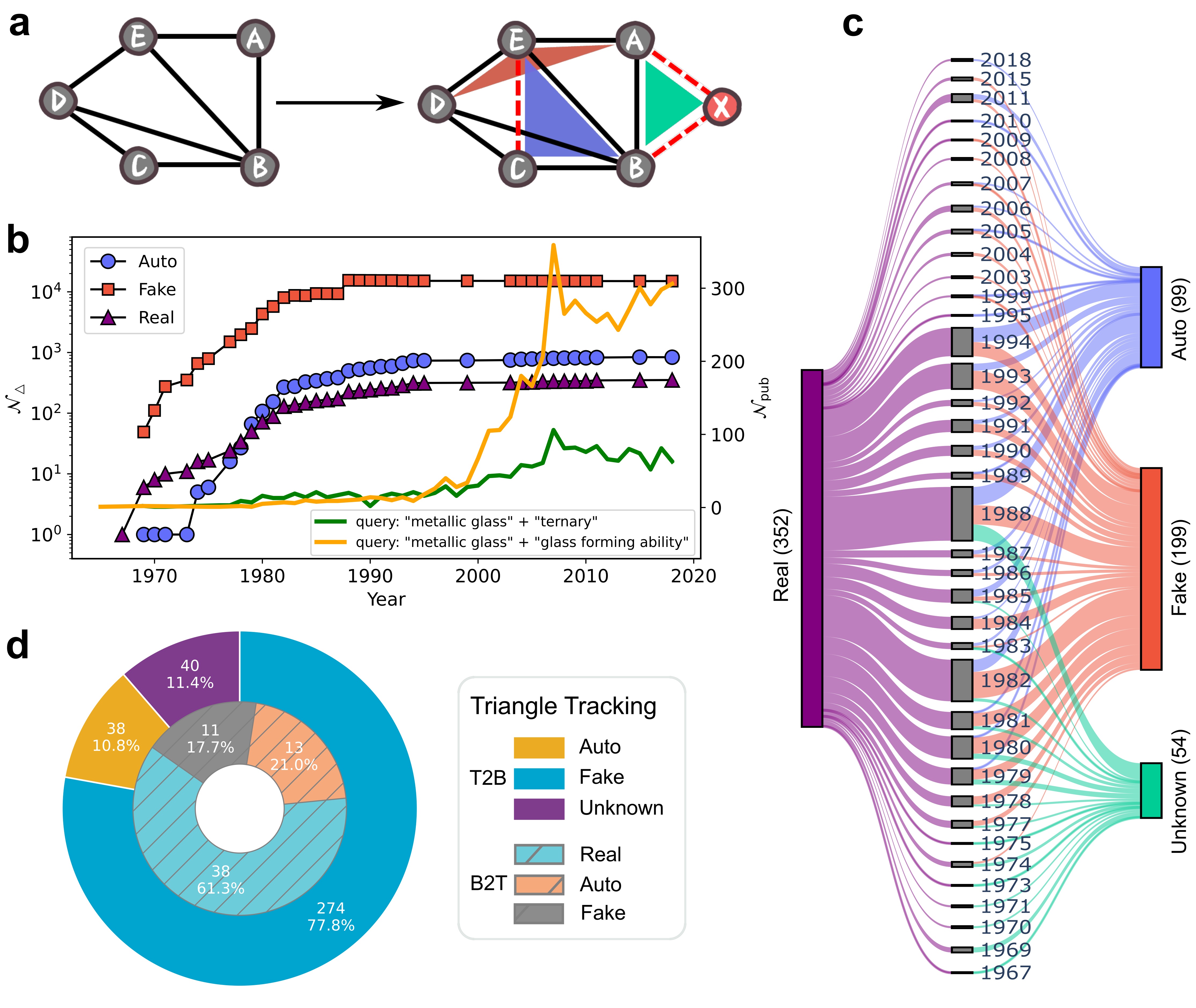}
	\caption{
		{\bf Triangle analysis in the ternary network.}
		(a) Schematic illustration of triangle classification. By adding new triangles ({\it CDE} and {\it ABX}, marked by the dashed red lines), there are three types of triangles in addition to the existing ones. The blue triangle ({\it BCE}) automatically forms (Auto). The red triangle ({\it ADE}) lacks an edge so it is a fake triangle (Fake). The green triangle is created by adding a new node so it is unknown from the left (Unknown).
		(b) The yearly growth of the number of classified triangles $\mathcal{N}_\Delta$. The number of publications $\mathcal{N}_{\rm pub}$ from the shown queries from the Web of Science database are imposed as lines for comparison.
		(c) A Sankey plot explaining the source of the developed ternary alloys (Real) at each year from the three groups.
		(d) Triangle tracking by cross-comparing the binary and ternary networks. The outer shows the source of the 352 developed ternaries from the binary network (T2B). The inner illustrates the source of the 62 automatically formed triangles in the binary network from the ternary network (B2T).
	}
	\label{fig3}
\end{figure}

In practice, ternary alloys usually have much better glass-forming ability than binary ones. That means larger sample sizes accessible from experiments, which is of great significance for property measurements and possible applications. This explains a more complicated network in Figs.~\ref{fig1}d, e than in Figs.~\ref{fig1}a, b. In fact, the ternary network is much harder in 3D printing. With this importance in mind, we carry out in-depth graph mining. We begin the discussion by defining several triangle units illustrated in Fig.~\ref{fig3}a. In the original schematic, there are five nodes with three triangle cliques, which represent the developed alloys. Starting with this topology, a new alloy, {\it CDE} is explored. It automatically generates a candidate triangle clique {\it BCE}, which is thus defined as an auto triangle (``Auto"). If not lacking a link, {\it ADE} would also become a clique, which is then defined as a fake triangle (``Fake"). These are the possible new glass-formers within the current network. In addition, by adding a new node X, there forms a triangle clique {\it ABX} that is unknown without X. So it is referred to as an unknown triangle (``Unknown"). The former falls into the innovation trap, while the latter escapes from it. Note that {\it ACE} belongs to Fake in the original plot.
More explicitly, in the ternary network in Fig.~\ref{fig1}d, there are 836 Auto triangles and 15,027 Fake triangles. If we look at the binary network in Fig.~\ref{fig1}a, there are 62 Auto and 8,374 Fake ones. 
(\blue{see Supplementary Materials for the lists of these triangles})
They may directly serve to guide experimental synthesis as a rule of thumb. 
\red{In fact, when we check the most recent discoveries (Co-Ni-Ta~\cite{HOWARD2025} and Al-Fe-Pr~\cite{LI2025}), they have been encoded in the ternary network (Auto for the former and Fake for the latter).}
\red{Incorporating molecular dynamics simulations and density-functional theory computations into the validation procedure is also of paramount significance in practice~\cite{tshitoyan_unsupervised_2019}.}

In Fig.~\ref{fig3}b, we explore the triangle entity breakdown in the dynamical material networks. In each year, we count the number of triangles falling into each class and show their numbers $\mathcal{N}_\Delta$ as a function of year. 
\blue{(see Supplementary Movies for the dynamical material networks)}
At first glance, there is rapid growth for all groups until approximately 1990. There are always many more Fake triangles in the network, suggesting the large material space. It is noteworthy to see that initially there were more Real triangles, which, nevertheless, intersected with that of Auto around 1980. This indicates the exploration efforts remaining from the network representation. The gaps among these groups remain until nowadays, because of the limited number of elements in amorphous alloy design. There is still a huge space to motivate newly guided experimental synthesis.
We also include the number of publications $\mathcal{N}_{\rm pub}$ queried from the Web of Science by using different keywords. We note an evolving curve that slowly grows when considering ``metallic glass" and ``ternary" as the query keywords. The growth almost stops after reaching its peak in $\sim2007$.
A rather different characteristic is observed when taking ``metallic glass" and ``glass forming ability" as the query keywords. In the period of intensive alloy discovery, there are publications that barely discuss the concept of glass-forming ability. The main research focus would be trying to create new amorphous alloys by trial-and-error experiments. After 2000, there is a quick climb in these publications, until saturation just before 2010.
This growth pattern marks the emergence of empirical rules and theoretical proposals on the prediction of glass-forming ability in the 2000s~\cite{takeuchi_classification_2005}. By summarizing the accumulated research experience and data before 2000, the field started to figure out how to design new materials with a better chance.

To unveil the innovation origin of the Real group, we show a Sankey plot in Fig.~\ref{fig3}c. It shows when the 352 ternary alloys were invented: mostly in the 1980s and 1990s. Meaningfully, they are broken into the groups of Unknown, Fake, and Auto. Only a very small portion of them is from Unknown, which always shows up in the early stage. The other two groups span over the past four decades. Strikingly, most of them are from Fake. The source from Auto indicates the effective design from the material network. Nevertheless, the success of Fake and Auto strongly recommends the prediction power of the ternary network.
\red{Since 1989, all the developed MGs were already encoded in the network. For example, new ternary MGs discovered in 1994 all came from the material network constructed from the ternary MGs developed before the year (Fake and Auto). This time-split validation strategy manifests the validation effectiveness in the past 20 years.} 
Mining the candidates from Fig.~\ref{fig3}c will be a valuable way to optimize the material design. 
\red{For instance, Co-Ni-Ta and Al-Fe-Pr, the latest experimental discoveries in 2025~\cite{HOWARD2025,LI2025}, have been encoded in the latest ternary network topology.}
\blue{(see Supplementary Materials for the lists of these groups)}

As a meaningful high-order object~\cite{millan_topology_2025}, the triangle entity naturally emerges in any network. We perform additional topology analysis to build hidden connections between the binary and ternary networks. In practice, we have two sets of triangles from these networks. By comparing them, we track the source of one group from the other. In the outer pie plot of Fig.~\ref{fig3}d, we unearth how the 352 ternary amorphous alloys could be designed from the binary network. In detail, we find that, remarkably, nearly $\sim 78\%$ of them are from the Fake triangles and $\sim 11\%$ from the Auto. Only the left is from the Unknown, demonstrating new elements in the ternary network.
In turn, we explore the prediction of the 62 Auto triangles in the binary network from the ternary one. Surprisingly, more than half of them actually form an amorphous alloy. Meanwhile, the rest are from the Auto and Fake groups, suggesting complete overlap between the two sets. 
This cross-comparison conveys important messages. The strong overlap indicates the hidden connection between amorphous alloys with mutual elements but different numbers of components. The magic minor alloying could be probably understood from this new viewpoint. Additionally, we may have the capability to predict multi-component amorphous alloys by exploring the space of much simpler ones. This will enormously reduce the complexity. This is one of our most important ongoing research efforts assisted by advanced AI techniques.

\red{Empirically, the glass-forming ability is usually superior if more components were involved. Simultaneously, the material space will grow logarithmically (see, for example, Fig. 1f). This renders limited available data for multi-component amorphous alloys. From the above cross-validations on the binary and ternary networks, in addition to explore the vast space of ternary alloys, the material networks are capable of predicting alloys with more components. For example, the cliques with $\mathcal{S}>3$ can serve as such material candidates for further screening. The network representations provide promising alternative routes to solve the glass-forming ability issue of metallic alloys with varying components.}

\begin{figure}[!t] 
	\centering
	\includegraphics[width = \textwidth]{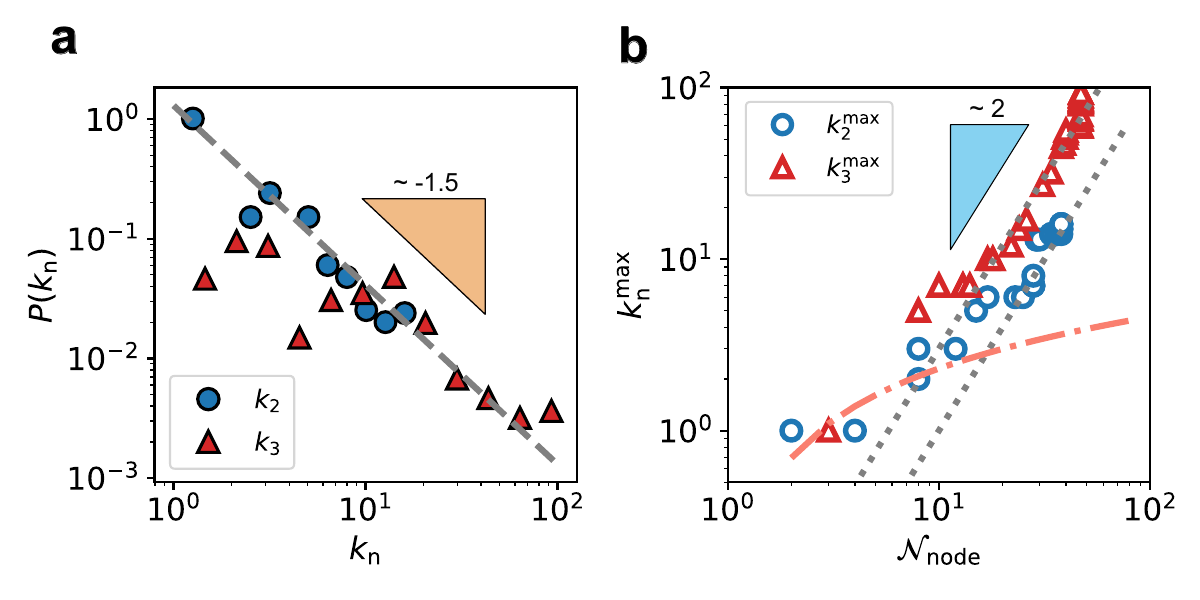}
	\caption{
		{\bf Scaling properties of the material networks.}
		(a) Probability distribution of the network entity. We consider the degree of nodes ($k_2$) and the number of triangles of nodes ($k_3$) in the binary and ternary networks, respectively. They are captured by a pow-law relation, $P(k_n)\sim k_n^{-\gamma}$ with $\gamma \approx 1.5$ (the dashed line). 
		(b) The dependence of the maximum of $k_n$ on the number of nodes $\mathcal{N}_{\rm node}$ from the dynamical networks (see Fig.~\ref{fig2}).
		The dotted lines represent a power-law scaling, $k_{n}^{\rm max} \sim \mathcal{N}_{\rm node}^{1/(\gamma-1)}$, where $\gamma \approx 1.5$ for both. 
		The orange dot-dashed line represents a logarithmic scaling for a random network ($k_{n}^{\rm max} \sim \ln \mathcal{N}_{\rm node}$).
	}
	\label{fig4}
\end{figure}

\subsection*{Analogies of material networks in daily lives}

One of the most crucial properties of a given graph is the degree distribution of nodes. The discovery of the scaling-free feature from actor collaboration graph, World Wide Web, and power grid opened up a new era of network science~\cite{barabasi_emergence_1999}. It provides fundamental classifications of graphs. The preferential attachment mechanism of network growth unveils the interesting trackable path of an evolving network, i.e. network dynamics. This has critical implications in various fields, for example, the Internet, social networks, biological networks, pandemics, etc.

We show the probability distributions of the network entities in Fig.~\ref{fig4}a. We emphasize that we consider the degree of nodes ($k_2$) for the binary network and 
the number of Real triangles ($k_3$) of nodes for the ternary network. 
A scaling behavior of $P(k_n) \sim k_n^{-\gamma}$ with $\gamma \approx 1.5$ is observed. 
The exponent $\gamma$ does not fall into the normal scale-free regime ($\gamma \in [2,3]$). The Barab$\acute{a}$si-Albert model provides a foundational mechanism to generate scale-free networks with $\gamma=3$~\cite{barabasi_emergence_1999}. Preferential attachment in the network growth is its most critical feature.
The scale-free feature indicates a very small chance for a node with an extremely large degree but does suggest the existence of hubs. The hub refers to a node with much higher degrees than others in a network. During growth, new nodes preferentially attach to the hubs, resulting in graph communities, a typical feature in large graphs. 

From the aforementioned analyses, these features are preserved in our networks. On the one hand, the network is growing with different probabilities for each node, which is driven by physical knowledge. On the other hand, there are ``small" elemental hubs, like Al, Fe, Ni, and Zr (see Fig.~\ref{fig1}h), which show preferential attachment features during alloy development. 
That is, we usually prefer to design an amorphous alloy with these elements involved.
Therefore, we classify our material networks into the abnormal scale-free class. This intrinsically suffers from the physical constraints of the limited number of elements in the periodic table. Even though all elements are supposed to be capable of being components in a metallic glass-former, the maximum number of nodes is 118. The enrichment of links or triangles will require lots of experimental effort. But at this stage, even the high-throughput experiments are incapable of sampling the material space efficiently, let alone the traditional method.

To distinguish the material networks from random networks, we compare the maximum degree $k_n^{\rm max}$ to the number of nodes $\mathcal{N}_{\rm node}$ in Fig.~\ref{fig4}b. To complement the scarcity of the material data, here we consider the dynamical networks for analysis. The outstanding observation is the faster growth of $k_n^{\rm max}$ following $\mathcal{N}_{\rm node}^{1/(\gamma-1)}$ with $\gamma \approx 1.5$ than the random growth following $k_n^{\rm max} \approx \ln \mathcal{N}_{\rm node}$. The increased discrepancy especially at large $\mathcal{N}_{\rm node}$ also indicates preferential attachment in the network growth.

\begin{figure}[!t] 
	\centering
	\includegraphics[width = \textwidth]{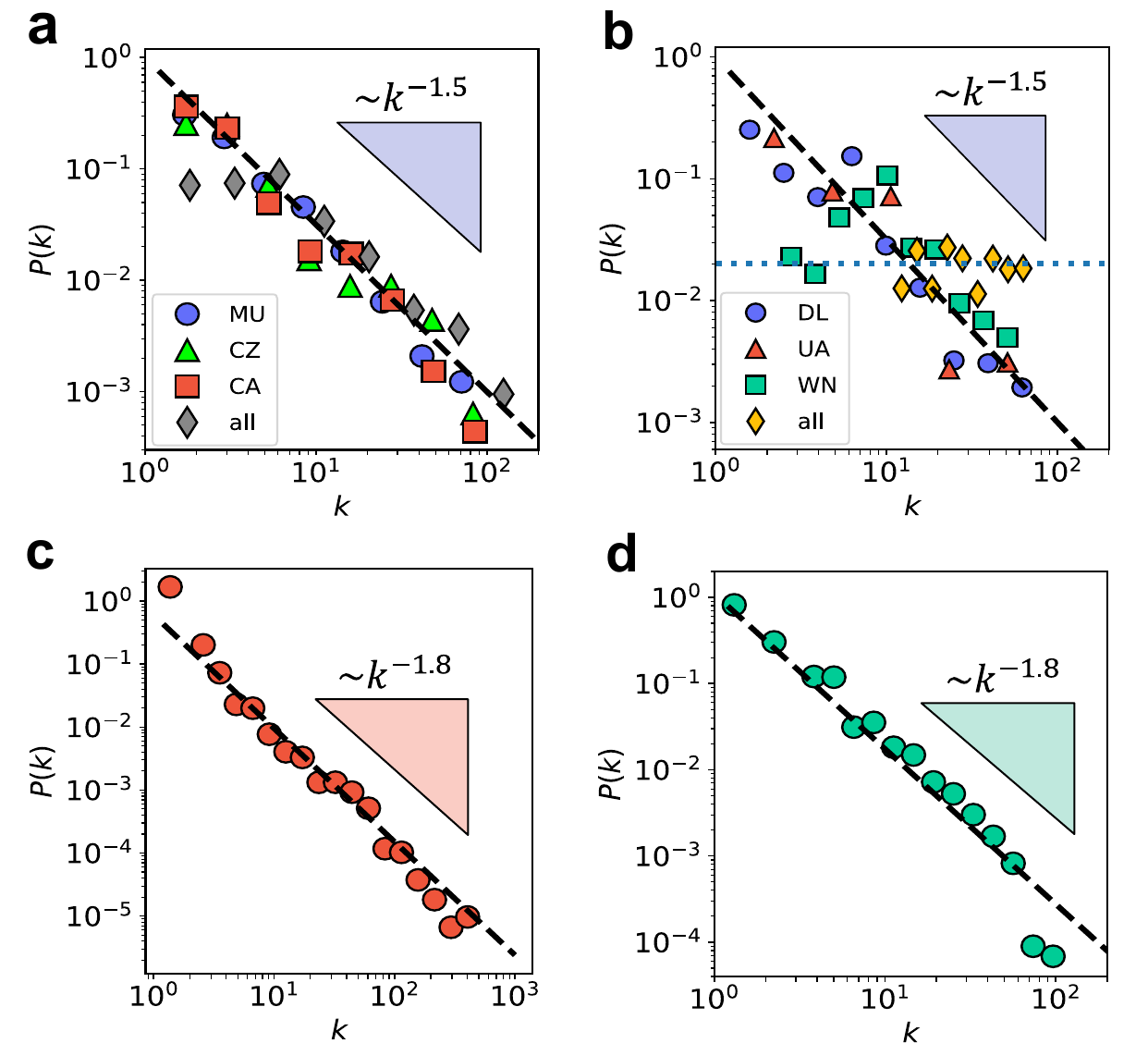}
	\caption{
		{\bf Scaling properties of the real-world networks.}
		The probability distributions of these networks are fitted to the power-law decay behaviour.
		(a) The Chinese flight network. The subgroup data for China Eastern Airlines (MU), China Southern Airlines (CZ), and Air China (CA) are shown for comparison.
		(b) The US flight network. The subgroup data for Delta Airlines (DL), United Airlines (UA), and Southwest Airlines (WN) are shown for comparison.
		\red{(c)} The email communication network. 
		\red{(d)} The blog network. 
		The power-law scaling, $P(k_n)\sim k_n^{-\gamma}$, holds for all these cases with $\gamma \approx 1.5$ for (a) and (b), and $\gamma \approx 1.8$ for (c) and (d). 
	}
	\label{fig5}
\end{figure}

We argue that the physical constraints-induced characteristics shall not be limited to our small physics-driven material networks. In reality, the networks with physical objects as nodes and their meaningful connections should show similar behaviors. It is definitely not true that the number of nodes and the corresponding node degrees can be extrapolated to infinity. However, different networks do behave at different length scales.
Motivated by this point, we explored various real-world networks and showed some typical examples in Fig.~\ref{fig5}, bearing the data availability.
The four examples are flight networks in China and in the US, email network, and blog network. In particular, the flight networks are broken down into different airlines. In all these cases, they faithfully obey a power-law decay with $P(k) \sim k^{-\gamma}$ with $\gamma < 2$, consistent with our material networks. This small exponent implies faster growth of links than nodes. The physical constraints in the flight networks are the airports and in the other two networks are the users.

\red{There are several possible factors that may affect the exponent. 
	Firstly, we note that $\gamma$ is slightly lower in the smaller networks (material  and flight networks) than the larger ones (email and blog networks). This indicates that large networks may suffer less from physical constraints as there are many more nodes available and thus many more ways to grow.  
	Secondly, we identified similar $\gamma$ from binary and ternary networks by focusing on the edge and the triangle, respectively. This indicates the effectiveness of the abnormal behavior for different effective edge definition. As our current work is pioneering in defining high-order objects as effective edges, there is plenty of room to quantitatively explore how they may influence the scaling behavior. 
	Thirdly, from our above discussion, adding non-existing edges (or missing data) will create stronger abnormality. Because this will increase the probability at higher node degrees (see Fig. 4a and Fig. 5). Selectively pruning over the networks may help reveal the hidden influence.}

\red{In Fig.~\ref{fig5}b, if looking separately for major airlines, the behavior, $p(k) \sim k^{-1.5}$, preserves. However, if we look at the combined data from all airlines, $p(k)$ shows a plateau in a small range of $k$. However, with 70 nodes available, the network will require 2415 undirected edges (or flights) to be fully connected to connect all pair nodes (cities). However, the flight network only has 1262 undirected edges, which guarantees sufficient traffic flow and gives $p(k)$ plateau.}
This may indicate that a fully connected network is not always necessary for optimal data flow.
\red{This scenario may apply to the material networks. For example, for the current ternary network, there are 47 nodes, which gives 16215 ternary candidates if fully connected. Currently, the number is 352, far less than the hypothetical value. 
	We thus speculate that when a sufficient number of triangles (much less than 16215) created, we may get enough knowledge to understand the network. Then, we may get direct suggestions for many other candidates that are not generated. Nevertheless, how to generate such sufficient triangles in a proper order (the optimal pathway) is very challenging. Designing this pathway will require comprehensive domain knowledge, such as amorphous alloys and network science. Once the sequence of alloy development is unveiled, it will be feasible to design targeted amorphous alloys in an efficient way. This deserves careful future investigations.}

This will bridge network studies from different fields and enable knowledge share. Our material networks strongly depend on our understanding of the fundamental physics at the atomic scale. However, the real-world networks will rely on large physical objects like airports or virtual human interactions. Apparently, they are distinct from each other. But from the perspective of networks with physical constraints, they show very similar properties, highlighting universal principles governing constrained growth in complex networks. 
This is especially significant in materials science as there is only limited number of elements available in the periodic table.
By thoroughly investigating the small networks, we may be able to understand the networks at a larger length scale much better. A good strategy will be considering the small network as the hierarchically coarse-grained mappings of large networks.

\vspace{1cm}
\section*{DISCUSSION}\label{sec3}

In this study, we propose the material networks for amorphous alloys. We focus on the binary and ternary metallic glass-formers developed in experiments since the birth of the field. 
The network representations are superior than the traditional tabular data representation in many aspects.
\red{It provides alternative strategies to deal with small datasets, which is commonly encountered in materials science research. Other than supervised machine learnings suitable for tabular representation, the networks are more easily integrated with deep learning models, such as graph neural networks.}

With the assistance of 3D printing, we effectively performed graph mining over the data-driven material networks. We also reveal the history of material discovery from the dynamical material networks. It demonstrates the innovation trap in materials discovery that the invented materials were already encoded in the material networks. That is, the material networks are capable of providing useful recommendations for new alloy design.
This is further corroborated by the topology analysis of the ternary network. This implies an efficient design strategy to take advantage of the network representations. 
There are large unexplored material spaces revealed by the material networks for binary and ternary amorphous alloys. We provided comprehensive recommendation lists for potential amorphous alloys from the network topology analysis.
\red{An alternative efficient way to make refined predictions or benchmarks is by advanced machine learning techniques. We can build recommendation systems by graph neural networks with versatile algorithms and architectures. The predictive power of these models has been witnessed and will be reported separately.}

In addition, the hidden connection between the binary and ternary networks indicates the capability of the network representation to make suggestions for multi-component amorphous alloys. These findings provide fresh ideas on optimal alloy design in the future.
More significantly, this network representation will play a crucial role in applying advanced artificial techniques in amorphous alloys design.

\red{In the current networks, we take the coarse-grained strategy by avoiding the compositions to reduce the complexity. In the vast material space, we noticed that it is quite challenging to directly pinpoint an exact composition. We thus take a detour to follow the top-down strategy: first identify a composition-free candidate system and then explore the phase diagram of the predicted system. A practical route for the latter is by high-throughput experiments (see, for example, ref.~\cite{li_high_2019}).
	In practice, for the binary network, we can integrate the compositions to the edge features; for the ternary network, we can integrate the compositions to either the involved edges or the triangle plane. 
	In addition, we can design weights for edges, triangles, or high-order objects, to reflect the properties of interest, such as maximum glass-forming ability, maximum critical size,  width of the glass-forming composition range, mechanical, chemical or physical properties.
	We can then design graph neural networks to incorporate these features to the model architecture to make composition-resolved predictions with desired material properties. This is our ongoing effort to tackle the grand challenge of amorphous alloy design. 
	The flexibility of integrating material properties into the network representations opens a new avenue to design more properties in an intelligent way.
	In addition, including high-order object features is an important research topic in graph neural networks. We may expect cross-disciplinary breakthroughs in the future.}

From the scaling analysis of the network degree of nodes defined from ordinary links and high-order triangles, we classify the material networks into the abnormal scale-free class. Nevertheless, it shows typical features deviating from random networks, for example, element hubs and network growth with preferential attachment. This inherently originates from the physical constraints, since there are a limited number of elements in the periodic table. The growth rate of link or triangle is much faster than that of the node.
We find analogous real-world networks that suffer from physical constraints, such as the fight networks, email communication network, and blog network. 
This similarity is likely to facilitate multi-disciplinary collaborative research and provide consulting clues for better decision-making. 
It also paves new ways to unveil the mystery of extremely large networks and to accelerate material discovery in a smarter way, especially in the era of AI.

\red{Looking foward, the advent of AI techniques motivates text processing to create insightful knowledge graphs for materials science~\cite{MRDJENOVICH2020,statt_rohr_guevarra_breeden_suram_gregoire_2023,venugopal_matkg_2024,ye2025constructionapplicationmaterialsknowledge}. They reveal the complex semantic relationships between various material entities, such as name, properties, synthesis methods, characterizations, applications, etc. The graph representation of these entities brings unprecedented insights hidden by the traditional plain tabular representation. From the sense of graph representation, our material networks share a lot of similarities with knowledge graphs. While knowledge graphs bring general semantic information, our material networks summarize the amorphous alloys reported in the literature and provide focused information on alloy development. It captures the specific relationship among elements to deliver sophisticated insights for new alloy development. This is beyond the capability of knowledge graph currently. However, careful distillation or refinement of comprehensive knowledge graphs will provide complement power to empower the material networks to make advanced predictions. Meanwhile, applying our strategies in the material networks, like dynamical networks, in designing knowledge graphs will also be beneficial. This is interesting for detailed future investigation to integrate the two fields.}

\vspace{1cm}
\section*{Methods}

\subsection*{Databases of amorphous alloys}
We collect the experimental data for the developed binary and ternary amorphous alloys reported in the literature. We primarily combine  the four databases from the previous works that are relevant to amorphous alloy research: 
(1) the dataset used for machine learning prediction of ternary alloys to form glasses from Ren et al.~\cite{ren_accelerated_2018}; 
(2) the dataset used to predict the glass-forming ability and supercooled liquid range of bulk amorphous alloys by Logan et al.~\cite{ward_machine_2018};
(3) the dataset provided by Schultz et al.~\cite{schultz_exploration_2021} that collects different physical properties for different amorphous alloys.
(4) the dataset of binary amorphous alloys created by Li et al. to explore the potential amorphous alloys~\cite{li_how_2017}.
To unify these datasets, we remove the redundant items and further clean the data to keep the unique records. Finally, there are 12,206 data in total reported so far, with each composition reported to be able to form a glass or not. We query the binary or ternary alloys with at least one composition reported as amorphous to make separate datasets. 
To track the development of these amorphous alloys, we also carefully checked these systems in the literature to extract the earliest year when a system is reported. We then consolidate these two datasets by neglecting the material compositions for further analysis.

\subsection*{Material network of binary amorphous alloys}
There are 94 binary systems reported so far in our database to have at least one composition of each system that has been fabricated into an amorphous alloy, no matter in the format of thin film, ribbon, or bulk. They are not differentiated here. Because binary systems usually have poor glass-forming ability and the available experimental data is scarce. We propose a graph representation of these systems by treating the involved elements as nodes and the binary systems as edges. Thus, a regular graph as shown in Fig.~\ref{fig1}a is constructed. To inspire deeper insights, we build this graph as a real three-dimensional network by 3D printing (see Fig.~\ref{fig1}b). The dynamical network is analyzed to acquire key scientific knowledge.

\subsection*{Material network of ternary amorphous alloys}
There are in total 352 ternary systems reported so far in our database to have at least one composition that has been fabricated into an amorphous alloy. We construct a graph representation for these systems by considering a higher-order unit, i.e. triangles,  rather than the links between pairs. By generating the three-dimensional network layout in Fig.~\ref{fig1}d by the Fruchterman-Reingold algorithm~\cite{Fruchterman_graph_1991}, there are 348 edges that connect any two elements as nodes. This structure is then printed to real-world objects by 3D printing to help gain better insights (see Fig.~\ref{fig1}e). We then perform comprehensive data mining over the dynamical networks for ternary alloys.

\subsection*{Material network construction algorithm}
The spatial layout of the material networks is generated using the Fruchterman-Reingold algorithm~\cite{Fruchterman_graph_1991}. It is a force-directed graph topology-built method that simulates a physical system.
The algorithm conceptualizes nodes as charged particles repelling each other, while edges behave as springs attracting connected nodes. The pair repulsive force and attracting force are defined as $f_{\text{rep}}(R) = -\frac{p^2}{R}$ and
$f_{\text{attr}}(R) = \frac{R^2}{p}$, respectively. $R$ denotes the Euclidean distance between nodes and $p$ is the optimal node separation constant. The final configuration is obtained by iteratively updating node positions based on the resultant forces until the system stabilizes, i.e. node displacements fall below a threshold. The network layout does not influence our current analyses.

\subsection*{Real-world networks}
To build hidden connections between scientific material networks and real-world instances, we came up with four analogous realistic networks that should suffer from physical constraints. These networks exist over different entities from totally different data resources.
The first one is the Chinese flight network which combines data from multiple domestic airlines, collected from public online sources. In total, there are 187 nodes (airports) and 1,761 edges (airlines). Note that different companies can fly the same route, but we do not consider this edge attribute in this study, similar to neglect compositions in the material networks.
The second one is the US airline flight network from major US carriers, developed using aviation datasets obtained from the OpenFlights database~\cite{openflights}. In total, there are 70 nodes and 1,262 undirected edges. 
The third network is the email communication network, sourcing from ref.~\cite{networkrepo}. There are 1,891 nodes and 4,465 undirected edges. Each node dictates an email account and each link demonstrates an existing email communication between the users.
The fourth is the blog network sourcing from refs.~\cite{Polblog_2005, networkrepo}. There are 663 nodes and 2,280 undirected edges. Each node means a blog and each edge indicates citation existence between blogs.
In our analysis, we focus on the scaling properties of the degree distribution of the nodes in these networks.

\vspace{1cm}

\begin{thebibliography}{10}
\expandafter\ifx\csname url\endcsname\relax
  \def\url#1{\texttt{#1}}\fi
\expandafter\ifx\csname urlprefix\endcsname\relax\def\urlprefix{URL }\fi
\providecommand{\bibinfo}[2]{#2}
\providecommand{\eprint}[2][]{\url{#2}}

\bibitem{de_materials_2021}
\bibinfo{author}{De~Leon, N.~P.}, \bibinfo{author}{Itoh, K.~M.},
  \bibinfo{author}{Kim, D.}, \bibinfo{author}{Mehta, K.~K.},
  \bibinfo{author}{Northup, T.~E.}, \bibinfo{author}{Paik, H.},
  \bibinfo{author}{Palmer, B.}, \bibinfo{author}{Samarth, N.},
  \bibinfo{author}{Sangtawesin, S.} \& \bibinfo{author}{Steuerman, D.~W.}
\newblock \bibinfo{title}{Materials challenges and opportunities for quantum
  computing hardware}.
\newblock \emph{\bibinfo{journal}{Science}} \textbf{\bibinfo{volume}{372}},
  \bibinfo{pages}{eabb2823} (\bibinfo{year}{2021}).

\bibitem{merchant_scaling_2023}
\bibinfo{author}{Merchant, A.}, \bibinfo{author}{Batzner, S.},
  \bibinfo{author}{Schoenholz, S.~S.}, \bibinfo{author}{Aykol, M.},
  \bibinfo{author}{Cheon, G.} \& \bibinfo{author}{Cubuk, E.~D.}
\newblock \bibinfo{title}{Scaling deep learning for materials discovery}.
\newblock \emph{\bibinfo{journal}{Nature}} \textbf{\bibinfo{volume}{624}},
  \bibinfo{pages}{80--85} (\bibinfo{year}{2023}).

\bibitem{li_how_2017}
\bibinfo{author}{Li, Y.}, \bibinfo{author}{Zhao, S.}, \bibinfo{author}{Liu,
  Y.}, \bibinfo{author}{Gong, P.} \& \bibinfo{author}{Schroers, J.}
\newblock \bibinfo{title}{How many bulk metallic glasses are there?}
\newblock \emph{\bibinfo{journal}{ACS Comb. Sci.}}
  \textbf{\bibinfo{volume}{19}}, \bibinfo{pages}{687--693}
  (\bibinfo{year}{2017}).

\bibitem{ding_combinatorial_2014}
\bibinfo{author}{Ding, S.}, \bibinfo{author}{Liu, Y.}, \bibinfo{author}{Li,
  Y.}, \bibinfo{author}{Liu, Z.}, \bibinfo{author}{Sohn, S.},
  \bibinfo{author}{Walker, F.~J.} \& \bibinfo{author}{Schroers, J.}
\newblock \bibinfo{title}{Combinatorial development of bulk metallic glasses}.
\newblock \emph{\bibinfo{journal}{Nat. Mater.}} \textbf{\bibinfo{volume}{13}},
  \bibinfo{pages}{494--500} (\bibinfo{year}{2014}).

\bibitem{li_high_2019}
\bibinfo{author}{Li, M.-X.}, \bibinfo{author}{Zhao, S.-F.},
  \bibinfo{author}{Lu, Z.}, \bibinfo{author}{Hirata, A.}, \bibinfo{author}{Wen,
  P.}, \bibinfo{author}{Bai, H.-Y.}, \bibinfo{author}{Chen, M.},
  \bibinfo{author}{Schroers, J.}, \bibinfo{author}{Liu, Y.} \&
  \bibinfo{author}{Wang, W.-H.}
\newblock \bibinfo{title}{High-temperature bulk metallic glasses developed by
  combinatorial methods}.
\newblock \emph{\bibinfo{journal}{Nature}} \textbf{\bibinfo{volume}{569}},
  \bibinfo{pages}{99--103} (\bibinfo{year}{2019}).

\bibitem{li_data_2022}
\bibinfo{author}{Li, M.-X.}, \bibinfo{author}{Sun, Y.-T.},
  \bibinfo{author}{Wang, C.}, \bibinfo{author}{Hu, L.-W.},
  \bibinfo{author}{Sohn, S.}, \bibinfo{author}{Schroers, J.},
  \bibinfo{author}{Wang, W.-H.} \& \bibinfo{author}{Liu, Y.-H.}
\newblock \bibinfo{title}{Data-driven discovery of a universal indicator for
  metallic glass forming ability}.
\newblock \emph{\bibinfo{journal}{Nat. Mater.}} \textbf{\bibinfo{volume}{21}},
  \bibinfo{pages}{165--172} (\bibinfo{year}{2022}).

\bibitem{jain_commentary_2013}
\bibinfo{author}{Jain, A.} \emph{et~al.}
\newblock \bibinfo{title}{Commentary: {The} {Materials} {Project}: {A}
  materials genome approach to accelerating materials innovation}.
\newblock \emph{\bibinfo{journal}{APL Mater.}} \textbf{\bibinfo{volume}{1}},
  \bibinfo{pages}{011002} (\bibinfo{year}{2013}).

\bibitem{batra_emerging_2021}
\bibinfo{author}{Batra, R.}, \bibinfo{author}{Song, L.} \&
  \bibinfo{author}{Ramprasad, R.}
\newblock \bibinfo{title}{Emerging materials intelligence ecosystems propelled
  by machine learning}.
\newblock \emph{\bibinfo{journal}{Nat. Rev. Mater.}}
  \textbf{\bibinfo{volume}{6}}, \bibinfo{pages}{655--678}
  (\bibinfo{year}{2021}).

\bibitem{curtarolo_NM_2013}
\bibinfo{author}{Curtarolo, S.}, \bibinfo{author}{Hart, G. L.~W.},
  \bibinfo{author}{Nardelli, M.~B.}, \bibinfo{author}{Mingo, N.},
  \bibinfo{author}{Sanvito, S.} \& \bibinfo{author}{Levy, O.}
\newblock \bibinfo{title}{The high-throughput highway to computational
  materials design}.
\newblock \emph{\bibinfo{journal}{Nat. Mater.}} \textbf{\bibinfo{volume}{12}},
  \bibinfo{pages}{191--201} (\bibinfo{year}{2013}).

\bibitem{szymanski_autonomous_2023}
\bibinfo{author}{Szymanski, N.~J.} \emph{et~al.}
\newblock \bibinfo{title}{An autonomous laboratory for the accelerated
  synthesis of novel materials}.
\newblock \emph{\bibinfo{journal}{Nature}} \textbf{\bibinfo{volume}{624}},
  \bibinfo{pages}{86--91} (\bibinfo{year}{2023}).

\bibitem{liu2024prompt}
\bibinfo{author}{Liu, S.}, \bibinfo{author}{Wen, T.},
  \bibinfo{author}{Pattamatta, A.~S.} \& \bibinfo{author}{Srolovitz, D.~J.}
\newblock \bibinfo{title}{A prompt-engineered large language model, deep
  learning workflow for materials classification}.
\newblock \emph{\bibinfo{journal}{Mater. Today}} \textbf{\bibinfo{volume}{80}},
  \bibinfo{pages}{240--249} (\bibinfo{year}{2024}).

\bibitem{bran_augment_2024}
\bibinfo{author}{Bran, A.~M.}, \bibinfo{author}{Cox, S.},
  \bibinfo{author}{Schilter, O.}, \bibinfo{author}{Baldassari, C.},
  \bibinfo{author}{White, A.~D.} \& \bibinfo{author}{Schwaller, P.}
\newblock \bibinfo{title}{Augmenting large language models with chemistry
  tools}.
\newblock \emph{\bibinfo{journal}{Nat. Mach. Intell.}}
  \textbf{\bibinfo{volume}{6}}, \bibinfo{pages}{525--535}
  (\bibinfo{year}{2024}).

\bibitem{klement1_non_1960}
\bibinfo{author}{Klement, W.}, \bibinfo{author}{Willens, R.} \&
  \bibinfo{author}{Duwez, P.}
\newblock \bibinfo{title}{Non-crystalline structure in solidified gold--silicon
  alloys}.
\newblock \emph{\bibinfo{journal}{Nature}} \textbf{\bibinfo{volume}{187}},
  \bibinfo{pages}{869--870} (\bibinfo{year}{1960}).

\bibitem{ward_general-purpose_2016}
\bibinfo{author}{Ward, L.}, \bibinfo{author}{Agrawal, A.},
  \bibinfo{author}{Choudhary, A.} \& \bibinfo{author}{Wolverton, C.}
\newblock \bibinfo{title}{A general-purpose machine learning framework for
  predicting properties of inorganic materials}.
\newblock \emph{\bibinfo{journal}{Npj Comput. Mater.}}
  \textbf{\bibinfo{volume}{2}}, \bibinfo{pages}{16028} (\bibinfo{year}{2016}).

\bibitem{takeuchi_classification_2005}
\bibinfo{author}{Takeuchi, A.} \& \bibinfo{author}{Inoue, A.}
\newblock \bibinfo{title}{Classification of bulk metallic glasses by atomic
  size difference, heat of mixing and period of constituent elements and its
  application to characterization of the main alloying element}.
\newblock \emph{\bibinfo{journal}{Mater. Trans.}}
  \textbf{\bibinfo{volume}{46}}, \bibinfo{pages}{2817--2829}
  (\bibinfo{year}{2005}).

\bibitem{sun2017machine}
\bibinfo{author}{Sun, Y.-T.}, \bibinfo{author}{Bai, H.-Y.},
  \bibinfo{author}{Li, M.-Z.} \& \bibinfo{author}{Wang, W.-H.}
\newblock \bibinfo{title}{Machine learning approach for prediction and
  understanding of glass-forming ability}.
\newblock \emph{\bibinfo{journal}{J. Phys. Chem. Lett.}}
  \textbf{\bibinfo{volume}{8}}, \bibinfo{pages}{3434--3439}
  (\bibinfo{year}{2017}).

\bibitem{xiong2020machine}
\bibinfo{author}{Xiong, J.}, \bibinfo{author}{Shi, S.-Q.} \&
  \bibinfo{author}{Zhang, T.-Y.}
\newblock \bibinfo{title}{A machine-learning approach to predicting and
  understanding the properties of amorphous metallic alloys}.
\newblock \emph{\bibinfo{journal}{Mater. Des.}} \textbf{\bibinfo{volume}{187}},
  \bibinfo{pages}{108378} (\bibinfo{year}{2020}).

\bibitem{liu2020machine}
\bibinfo{author}{Liu, X.}, \bibinfo{author}{Li, X.}, \bibinfo{author}{He, Q.},
  \bibinfo{author}{Liang, D.}, \bibinfo{author}{Zhou, Z.}, \bibinfo{author}{Ma,
  J.}, \bibinfo{author}{Yang, Y.} \& \bibinfo{author}{Shen, J.}
\newblock \bibinfo{title}{Machine learning-based glass formation prediction in
  multicomponent alloys}.
\newblock \emph{\bibinfo{journal}{Acta Mater.}} \textbf{\bibinfo{volume}{201}},
  \bibinfo{pages}{182--190} (\bibinfo{year}{2020}).

\bibitem{zhou_rational_2021}
\bibinfo{author}{Zhou, Z.}, \bibinfo{author}{He, Q.}, \bibinfo{author}{Liu,
  X.}, \bibinfo{author}{Wang, Q.}, \bibinfo{author}{Luan, J.},
  \bibinfo{author}{Liu, C.} \& \bibinfo{author}{Yang, Y.}
\newblock \bibinfo{title}{Rational design of chemically complex metallic
  glasses by hybrid modeling guided machine learning}.
\newblock \emph{\bibinfo{journal}{Npj Comput. Mater.}}
  \textbf{\bibinfo{volume}{7}}, \bibinfo{pages}{138} (\bibinfo{year}{2021}).

\bibitem{afflerbach_machine_2022}
\bibinfo{author}{Afflerbach, B.~T.} \emph{et~al.}
\newblock \bibinfo{title}{Machine learning prediction of the critical cooling
  rate for metallic glasses from expanded datasets and elemental features}.
\newblock \emph{\bibinfo{journal}{Chem. Mater.}} \textbf{\bibinfo{volume}{34}},
  \bibinfo{pages}{2945--2954} (\bibinfo{year}{2022}).

\bibitem{yao2022balancing}
\bibinfo{author}{Yao, Y.}, \bibinfo{author}{Sullivan~IV, T.},
  \bibinfo{author}{Yan, F.}, \bibinfo{author}{Gong, J.} \& \bibinfo{author}{Li,
  L.}
\newblock \bibinfo{title}{Balancing data for generalizable machine learning to
  predict glass-forming ability of ternary alloys}.
\newblock \emph{\bibinfo{journal}{Scr. Mater.}} \textbf{\bibinfo{volume}{209}},
  \bibinfo{pages}{114366} (\bibinfo{year}{2022}).

\bibitem{liu2023machine}
\bibinfo{author}{Liu, G.}, \bibinfo{author}{Sohn, S.}, \bibinfo{author}{Kube,
  S.~A.}, \bibinfo{author}{Raj, A.}, \bibinfo{author}{Mertz, A.},
  \bibinfo{author}{Nawano, A.}, \bibinfo{author}{Gilbert, A.},
  \bibinfo{author}{Shattuck, M.~D.}, \bibinfo{author}{O'Hern, C.~S.} \&
  \bibinfo{author}{Schroers, J.}
\newblock \bibinfo{title}{Machine learning versus human learning in predicting
  glass-forming ability of metallic glasses}.
\newblock \emph{\bibinfo{journal}{Acta Mater.}} \textbf{\bibinfo{volume}{243}},
  \bibinfo{pages}{118497} (\bibinfo{year}{2023}).

\bibitem{forrest2023evolutionary}
\bibinfo{author}{Forrest, R.~M.} \& \bibinfo{author}{Greer, A.~L.}
\newblock \bibinfo{title}{Evolutionary design of machine-learning-predicted
  bulk metallic glasses}.
\newblock \emph{\bibinfo{journal}{Digit. Discov.}}
  \textbf{\bibinfo{volume}{2}}, \bibinfo{pages}{202--218}
  (\bibinfo{year}{2023}).

\bibitem{zhou2023generative}
\bibinfo{author}{Zhou, Z.}, \bibinfo{author}{Shang, Y.}, \bibinfo{author}{Liu,
  X.} \& \bibinfo{author}{Yang, Y.}
\newblock \bibinfo{title}{A generative deep learning framework for inverse
  design of compositionally complex bulk metallic glasses}.
\newblock \emph{\bibinfo{journal}{Npj Comput. Mater.}}
  \textbf{\bibinfo{volume}{9}}, \bibinfo{pages}{15} (\bibinfo{year}{2023}).

\bibitem{liu2024effective}
\bibinfo{author}{Liu, G.}, \bibinfo{author}{Sohn, S.}, \bibinfo{author}{O'Hern,
  C.~S.}, \bibinfo{author}{Gilbert, A.~C.} \& \bibinfo{author}{Schroers, J.}
\newblock \bibinfo{title}{Effective subgrouping enhances machine learning
  prediction in complex materials science phenomena: Inoue's subgrouping in
  discovering bulk metallic glasses}.
\newblock \emph{\bibinfo{journal}{Acta Mater.}} \textbf{\bibinfo{volume}{265}},
  \bibinfo{pages}{119590} (\bibinfo{year}{2024}).

\bibitem{ren_accelerated_2018}
\bibinfo{author}{Ren, F.}, \bibinfo{author}{Ward, L.},
  \bibinfo{author}{Williams, T.}, \bibinfo{author}{Laws, K.~J.},
  \bibinfo{author}{Wolverton, C.}, \bibinfo{author}{Hattrick-Simpers, J.} \&
  \bibinfo{author}{Mehta, A.}
\newblock \bibinfo{title}{Accelerated discovery of metallic glasses through
  iteration of machine learning and high-throughput experiments}.
\newblock \emph{\bibinfo{journal}{Sci. Adv.}} \textbf{\bibinfo{volume}{4}},
  \bibinfo{pages}{eaaq1566} (\bibinfo{year}{2018}).

\bibitem{karniadakis_physics_2021}
\bibinfo{author}{Karniadakis, G.~E.}, \bibinfo{author}{Kevrekidis, I.~G.},
  \bibinfo{author}{Lu, L.}, \bibinfo{author}{Perdikaris, P.},
  \bibinfo{author}{Wang, S.} \& \bibinfo{author}{Yang, L.}
\newblock \bibinfo{title}{Physics-informed machine learning}.
\newblock \emph{\bibinfo{journal}{Nat. Rev. Phys.}}
  \textbf{\bibinfo{volume}{3}}, \bibinfo{pages}{422--440}
  (\bibinfo{year}{2021}).

\bibitem{ward_machine_2018}
\bibinfo{author}{Ward, L.}, \bibinfo{author}{O'Keeffe, S.~C.},
  \bibinfo{author}{Stevick, J.}, \bibinfo{author}{Jelbert, G.~R.},
  \bibinfo{author}{Aykol, M.} \& \bibinfo{author}{Wolverton, C.}
\newblock \bibinfo{title}{A machine learning approach for engineering bulk
  metallic glass alloys}.
\newblock \emph{\bibinfo{journal}{Acta Mater.}} \textbf{\bibinfo{volume}{159}},
  \bibinfo{pages}{102--111} (\bibinfo{year}{2018}).

\bibitem{schultz_exploration_2021}
\bibinfo{author}{Schultz, L.~E.}, \bibinfo{author}{Afflerbach, B.},
  \bibinfo{author}{Francis, C.}, \bibinfo{author}{Voyles, P.~M.},
  \bibinfo{author}{Szlufarska, I.} \& \bibinfo{author}{Morgan, D.}
\newblock \bibinfo{title}{Exploration of characteristic temperature
  contributions to metallic glass forming ability}.
\newblock \emph{\bibinfo{journal}{Comput. Mater. Sci.}}
  \textbf{\bibinfo{volume}{196}}, \bibinfo{pages}{110494}
  (\bibinfo{year}{2021}).

\bibitem{barabasi_emergence_1999}
\bibinfo{author}{Barab{\'a}si, A.-L.} \& \bibinfo{author}{Albert, R.}
\newblock \bibinfo{title}{Emergence of scaling in random networks}.
\newblock \emph{\bibinfo{journal}{Science}} \textbf{\bibinfo{volume}{286}},
  \bibinfo{pages}{509--512} (\bibinfo{year}{1999}).

\bibitem{millan_topology_2025}
\bibinfo{author}{Mill{\'a}n, A.~P.}, \bibinfo{author}{Sun, H.},
  \bibinfo{author}{Giambagli, L.}, \bibinfo{author}{Muolo, R.},
  \bibinfo{author}{Carletti, T.}, \bibinfo{author}{Torres, J.~J.},
  \bibinfo{author}{Radicchi, F.}, \bibinfo{author}{Kurths, J.} \&
  \bibinfo{author}{Bianconi, G.}
\newblock \bibinfo{title}{Topology shapes dynamics of higher-order networks}.
\newblock \emph{\bibinfo{journal}{Nat. Phys.}} \textbf{\bibinfo{volume}{21}},
  \bibinfo{pages}{353--361} (\bibinfo{year}{2025}).

\bibitem{boguna_network_2021}
\bibinfo{author}{Bogu{\~n}{\'a}, M.}, \bibinfo{author}{Bonamassa, I.},
  \bibinfo{author}{{De Domenico}, M.}, \bibinfo{author}{Havlin, S.},
  \bibinfo{author}{Krioukov, D.} \& \bibinfo{author}{Serrano, M.~{\'A}.}
\newblock \bibinfo{title}{Network geometry}.
\newblock \emph{\bibinfo{journal}{Nat. Rev. Phys.}}
  \textbf{\bibinfo{volume}{3}}, \bibinfo{pages}{114--135}
  (\bibinfo{year}{2021}).

\bibitem{albert_statistical_2002}
\bibinfo{author}{Albert, R.} \& \bibinfo{author}{Barab{\'a}si, A.-L.}
\newblock \bibinfo{title}{Statistical mechanics of complex networks}.
\newblock \emph{\bibinfo{journal}{Rev. Mod. Phys.}}
  \textbf{\bibinfo{volume}{74}}, \bibinfo{pages}{47} (\bibinfo{year}{2002}).

\bibitem{HOWARD2025}
\bibinfo{author}{Howard, J.}, \bibinfo{author}{Suenram, G.},
  \bibinfo{author}{Thompson, F.}, \bibinfo{author}{Murray, P.},
  \bibinfo{author}{Chidambaram, D.}, \bibinfo{author}{Crawford, G.} \&
  \bibinfo{author}{Carlson, K.}
\newblock \bibinfo{title}{Computational design and experimental verification of
  {Ta-Ni-Co} metallic glasses produced via gas atomization}.
\newblock \emph{\bibinfo{journal}{Acta Mater.}} \textbf{\bibinfo{volume}{296}},
  \bibinfo{pages}{121206} (\bibinfo{year}{2025}).

\bibitem{LI2025}
\bibinfo{author}{Li, Z.}, \bibinfo{author}{Yuan, J.}, \bibinfo{author}{Zhai,
  T.}, \bibinfo{author}{Wang, Q.}, \bibinfo{author}{Ding, D.} \&
  \bibinfo{author}{Xia, L.}
\newblock \bibinfo{title}{Synthesis of {Fe88Pr10Al2} metallic glass with
  excellent magnetocaloric effect near the ice point}.
\newblock \emph{\bibinfo{journal}{J. Non-Cryst. Solids}}
  \textbf{\bibinfo{volume}{662}}, \bibinfo{pages}{123576}
  (\bibinfo{year}{2025}).

\bibitem{tshitoyan_unsupervised_2019}
\bibinfo{author}{Tshitoyan, V.}, \bibinfo{author}{Dagdelen, J.},
  \bibinfo{author}{Weston, L.}, \bibinfo{author}{Dunn, A.},
  \bibinfo{author}{Rong, Z.}, \bibinfo{author}{Kononova, O.},
  \bibinfo{author}{Persson, K.~A.}, \bibinfo{author}{Ceder, G.} \&
  \bibinfo{author}{Jain, A.}
\newblock \bibinfo{title}{Unsupervised word embeddings capture latent knowledge
  from materials science literature}.
\newblock \emph{\bibinfo{journal}{Nature}} \textbf{\bibinfo{volume}{571}},
  \bibinfo{pages}{95--98} (\bibinfo{year}{2019}).

\bibitem{MRDJENOVICH2020}
\bibinfo{author}{Mrdjenovich, D.}, \bibinfo{author}{Horton, M.~K.},
  \bibinfo{author}{Montoya, J.~H.}, \bibinfo{author}{Legaspi, C.~M.},
  \bibinfo{author}{Dwaraknath, S.}, \bibinfo{author}{Tshitoyan, V.},
  \bibinfo{author}{Jain, A.} \& \bibinfo{author}{Persson, K.~A.}
\newblock \bibinfo{title}{propnet: A knowledge graph for materials science}.
\newblock \emph{\bibinfo{journal}{Matter}} \textbf{\bibinfo{volume}{2}},
  \bibinfo{pages}{464--480} (\bibinfo{year}{2020}).

\bibitem{statt_rohr_guevarra_breeden_suram_gregoire_2023}
\bibinfo{author}{Statt, M.~J.}, \bibinfo{author}{Rohr, B.~A.},
  \bibinfo{author}{Guevarra, D.}, \bibinfo{author}{Breeden, J.},
  \bibinfo{author}{Suram, S.~K.} \& \bibinfo{author}{Gregoire, J.~M.}
\newblock \bibinfo{title}{The materials experiment knowledge graph}.
\newblock \emph{\bibinfo{journal}{Digit. Discov.}}
  \textbf{\bibinfo{volume}{2}}, \bibinfo{pages}{909–914}
  (\bibinfo{year}{2023}).

\bibitem{venugopal_matkg_2024}
\bibinfo{author}{Venugopal, V.} \& \bibinfo{author}{Olivetti, E.}
\newblock \bibinfo{title}{{MatKG}: An autonomously generated knowledge graph in
  material science}.
\newblock \emph{\bibinfo{journal}{Sci. Data}} \textbf{\bibinfo{volume}{11}}
  (\bibinfo{year}{2024}).

\bibitem{ye2025constructionapplicationmaterialsknowledge}
\bibinfo{author}{Ye, Y.}, \bibinfo{author}{Ren, J.}, \bibinfo{author}{Wang,
  S.}, \bibinfo{author}{Wan, Y.}, \bibinfo{author}{Razzak, I.},
  \bibinfo{author}{Hoex, B.}, \bibinfo{author}{Wang, H.}, \bibinfo{author}{Xie,
  T.} \& \bibinfo{author}{Zhang, W.}
\newblock \bibinfo{title}{Construction and application of materials knowledge
  graph in multidisciplinary materials science via large language model}.
\newblock In \emph{\bibinfo{booktitle}{Proceedings of the 38th International
  Conference on Neural Information Processing Systems}}, NIPS '24,
  \bibinfo{pages}{56878--56897} (\bibinfo{year}{2025}).

\bibitem{Fruchterman_graph_1991}
\bibinfo{author}{Fruchterman, T.~M.} \& \bibinfo{author}{Reingold, E.~M.}
\newblock \bibinfo{title}{Graph drawing by force-directed placement}.
\newblock \emph{\bibinfo{journal}{Softw. Pract. Exp.}}
  \textbf{\bibinfo{volume}{21}}, \bibinfo{pages}{1129--1164}
  (\bibinfo{year}{1991}).

\bibitem{openflights}
\bibinfo{title}{{OpenFlights}}.
\newblock \bibinfo{howpublished}{Online at \url{https://openflights.org/}}.
\newblock \urlprefix\url{https://openflights.org/}.

\bibitem{networkrepo}
\bibinfo{author}{Rossi, R.~A.} \& \bibinfo{author}{Ahmed, N.~K.}
\newblock \bibinfo{title}{The network data repository with interactive graph
  analytics and visualization}.
\newblock In \emph{\bibinfo{booktitle}{AAAI}} (\bibinfo{year}{2015}).

\bibitem{Polblog_2005}
\bibinfo{author}{Adamic, L.~A.} \& \bibinfo{author}{Glance, N.}
\newblock \bibinfo{title}{The political blogosphere and the 2004 u.s. election:
  divided they blog}.
\newblock In \emph{\bibinfo{booktitle}{Proc. 3rd Int. Worksh. Link Discov.}},
  LinkKDD '05, \bibinfo{pages}{36–43} (\bibinfo{year}{2005}).

\end{thebibliography}

\subsection *{Acknowledgments.}
This work is supported by the National Natural Science Foundation of China (Grant No. 52471178).
The support from the Chinese Academy of Sciences (XDB0510000) is also acknowledged.

\subsection *{Author contributions.}
Y.C.H. conceived and supervised the project. Y.C.H., S.Y.Z. and J.T. collected, cleaned and processed the data. S.L.L. and H.M.Z. helped the 3D-printing. S.Y.Z. and Y.C.H. prepared the plots. All the authors contributed to the data analysis and results discussion. Y.C.H. wrote the manuscript.

\subsection *{Competing interests.}
The authors declare no competing interests.

\subsection *{Correspondence.}
Correspondence and requests for materials should be addressed to Y.C.H.

\balance
\end{document}